\documentstyle[epsf]{article}
\newcommand{\be}{\begin{equation}}
\newcommand{\ee}{\end{equation}}
\newcommand{\bea}{\begin{eqnarray}}
\newcommand{\eea}{\end{eqnarray}}
\newcommand{\beas}{\begin{eqnarray*}}
\newcommand{\eeas}{\end{eqnarray*}}
\newcommand{\s}{{\bf s}}
\newcommand{\se}{{\bf \tilde{s}}}
\newcommand{\bookfig}[5]{
\begin{figure}\centering\fbox{\epsfysize=#5cm \epsfbox{#1.eps}}
\caption[#2]{#4}\label{#3}
\end{figure}
}
\begin{document}
\title{Shuffling Quantum Field Theory\footnote{MZ-TH/99-55, to appear
in {\bf Lett.~Math.~Physics}}}
\author{D.~Kreimer\thanks{Heisenberg
Fellow, Mainz Univ., {\small
kreimer@thep.physik.uni-mainz.de}}\\{\small Lyman Lab., Harvard
Univ.}\\{\small  Cambridge, Mass.02138, USA} }
\date{December 1999, {\small hep-th/9912290}}
\maketitle
\begin{abstract}
We discuss shuffle identities between Feynman graphs using the
Hopf algebra structure of perturbative quantum field theory. For
concrete exposition, we discuss vertex function in massless Yukawa
theory.
\end{abstract}
\section{Introduction}
The perturbative expansion of a quantum field theory is given in
terms of Feynman graphs which demand integration over positions of
all internal vertices. Such spacetime integrations diverge due to
singularities in the integrand 
located in the regions where the positions of two or
more such vertices coincide. The required process of
renormalization amounts to a continuation of the generalized
functions provided by Feynman integrands. Those integrands are
well defined on the configuration space of vertices at distinct
locations and have to be continued to diagonals so that the
resulting analytic expressions are integrable along the diagonals.
Thus, morally speaking, renormalization is a problem dual to
compactification of configuration space, but generalizes such
compactifications by allowing for the freedom of scale variations
in this continuation, resulting in the presence of the
renormalization group. From this viewpoint, it is no surprise that
the common algebraic structure underlying any perturbative quantum
field theory is a Hopf algebra based on rooted trees, as the
latter are known to stratify the various limits to diagonals in
configuration space \cite{FulMcP}.

While the Hopf algebra of rooted trees is the universal object for
the Hopf algebra structure of perturbative quantum field theory
\cite{CK1}, for any specific theory under consideration we can
formulate the Hopf algebra directly on Feynman graphs
\cite{Koverl,CK4}. In this latter formulation, the primitive
generators of the Hopf algebra ${\cal H}$ are graphs free of
subdivergences. The only non-trivial part of this Hopf algebra is
spanned by the graphs with a non-vanishing overall degree of
divergence, which form a sub-Hopf algebra ${\cal H}_c$ whose
primitive generators are graphs which are overall divergent, but
have no subdivergences. Such graphs provide  well-defined
scheme-independent coefficients of divergence.

This algebraic set-up completely settles the renormalization
procedure as a mathematical well-defined principle, and dispenses
with all conceptual criticism of local point-particle quantum
field theory which is based on the appearance of ultra-violet
divergences \cite{CK1,Koverl,CK4,Khopf,Kchen,CK2,CK3}. No need
remains to abandon local quantum field theories for that reason:
the mathematics of the perturbation series of a local quantum
field theory is, as far as the UV sector is concerned, sound and,
to my mind, rich and beautiful.

With the mathematics of renormalization  laid out in
\cite{CK1,CK4,Khopf,Kchen} in particular, we can ask for more
structure which can be revealed thanks to the Hopf algebra
structure of renormalization.

For a long time, David Broadhurst and the author observed
interesting number-theoretic content in the evaluation of Feynman
graphs without subdivergences, hence precisely in the primitive
generators of the very Hopf algebra of graphs.

Intuitively, a close connection between the topology of such a
primitive graph and its coefficient of overall divergence was
found by relating the topology of the graphs to braid-positive
knots in a highly empirical fashion \cite{KNF}. Nevertheless, a
faithful knot-to-number dictionary was established up to seven
loops and more. Up to the six-loop level all coefficients of
divergence were found to be multiple zeta values, while at the
seven-loop level a few coefficients remained unidentified, and may
well be shown, eventually, to point towards unexpected
generalizations of these classes of numbers.\footnote{Our results
in \cite{KNF} do not reduce the number content of field theories
in even dimensions to multiple zeta values and thus do not yet
fully support the conjecture in \cite{maxim}.}

The algebraic structure of those numbers can be reasonably
expected to be mirrored by algebraic relations between Feynman
diagrams.

In this paper, we want to make first attempts to describe such
relations between diagrams. We will consider quasi-shuffle
identities, as they were considered for example in \cite{MEH},
which are known by many to form part of the algebraic structure of
Euler/Zagier sums \cite{EZ}.

We will formulate these quasi-shuffle identities making use of the
operators $B_+$ (the closed Hochschild one-cocycle which exists is
any Hopf algebra of the type considered here) and $B_-$, which is
its inverse on the linear basis of decorated rooted trees, but
fails to be its inverse on products of rooted trees:
$B_+B_-(\prod_i t_i)\not=0$, see below.

We can then naturally formulate the shuffle algebra on decorated
rooted trees, and will turn to massless Yukawa theory for concrete
exposition.

We show that the iteration of vertex corrections provides, modulo
finite terms, a quasi-shuffle algebra, where letters are provided
by all two-line irreducible four-fermion skeleton graphs.

We will finish the paper by showing that the term which determines
the deviation from a proper shuffle identity obeys a pentagon
identity in its finite part.
\section{Quasi shuffle identities}
Let $A=a_1,a_2,\ldots,$ be a locally finite graded set of letters,
and $W=k\langle A\rangle$ be the $k$-vectorspace of words built
from such letters, with $e$, the empty word, being the only word
of degree zero. For all letters $x,y\in A$ and words $w_1,w_2\in
W$ let the shuffle product $\s:W \times W\to W$ be defined by
$\s[e,w_1]=\s[w_1,e]=w_1$ and
\be
\s[xw_1,yw_2]=x\s[w_1,yw_2]+y\s[xw_1,w_2]. \ee Then, $(W,\s)$ is
known to form a commutative algebra (hence, $\s$ is commutative
and associative), the shuffle algebra.

If we further set $\se[e,w_1]=\se[w_1,e]=w_1$ and
\be
\se[xw_1,yw_2]=x\se[w_1,yw_2]+y\se[xw_1,w_2]+C[x,y]\se[w_1,w_2],
\ee then  $(W,\se)$ forms again a commutative associative algebra
provided $C:A\times A\to A$ is commutative, associative and adds
degrees. See \cite{MEH} for proof and details. Also, a proof of
this claim will be given below in terms of operators $B_+,B_-$. We
call $(W,\se)$ a quasi-shuffle algebra.

Consider an algebra homomorphism $\rho: (W,\s)\to V$ where $V$ is
some algebra, $\rho(\s[w_1,w_2])=\rho(w_1)\rho(w_2)$. A prominent
example is provided by iterated integrals \cite{Kassel}, where
letters $a_i\in A$ are represented by differential one-forms, and
the iterated integral $F^w(x)$ fulfills the  shuffle identity
$F^{\s(w_1,w_2)}(x)=F^{w_1}(x)F^{w_2}(x)$. Essentially, this
identity allows to rewrite any rooted tree with sidebranchings as
a sum over rooted trees without sidebranchings \cite{Kchen}.

Consider now decorated rooted trees \cite{CK1} where decorations
are taken from the alphabet $A$. Let ${\cal H}(A)$ be the
resulting Hopf algebra \cite{CK1,Koverl}. As usual, we have
operators $B_-,B_+^x$, $x\in A$, such that  $B_-(\prod_{i=1}^k
t_i)=\sum_{i=1}^k t_1\ldots B_-(t_i)\ldots t_k$ and we write
$B_+B_-$ for
\be
B_+B_-(\prod_{i=1}^k t_i):=\sum_{i=1}^k B_+^{r(t_i)}(t_1\ldots
B_-(t_i)\ldots t_k), \ee where $r(t)\in A$ gives the decoration at
the root. With these definitions the identity operator $id:H\to
H$, $id(X)=X$ can be written as $id= B_-B_+^x$ for all $x\in A$.

We can formulate a (quasi)-shuffle product on decorated rooted
trees as follows. We identify a $k$-letter word
$a_{i_1}a_{i_2}\ldots a_{i_k}$ with a decorated rooted tree
without sidebranchings, $$ T=B_+^{a_{i_1}}\circ
B_+^{a_{i_2}}\circ\ldots\circ B_+^{a_{i_k}}(e),$$ and write  the
shuffle product as \be
\s[t_1,t_2]=B_+^{r(t_1)}(\s[B_-(t_1),t_2])+B_+^{r(t_2)}(\s[t_1,B_-(t_2)]).
\ee This is well-defined as $B_-(t)$ is again a single rooted
tree, due to the fact that $T$ has no sidebranching. Let ${\cal
H}_1(A)$ be the subset in ${\cal H}(A)$ of single decorated rooted
trees, ie.~the linear basis of ${\cal H}(A)$. In general, on
${\cal H}_1(A)$, $B_-$ is a map ${\cal H}_1(A)\to {\cal H}(A)$
while on rooted trees without sidebranchings, $B_-$ maps ${\cal
H}_1(A)\to {\cal H}_1(A)$.

We can define the shuffle product for  arbitrary decorated rooted
trees $t\in{\cal H}_1(A)$ using  $u:{\cal H}(A)\to {\cal H}_1(A)$
given in an iterative manner by $u(\prod_{i=1}^k
t_i)=\s[t_1,u(\prod_{i=2}^k t_i)]$, $u(t_i)=t_i$:
\be
\s[t_1,t_2]=B_+^{r(t_1)}(\s[u(B_-(t_1)),t_2])+B_+^{r(t_2)}(\s[t_1,u(B_-(t_2))]).
\ee

We obtain a quasi-shuffle algebra on decorated rooted trees if we
replace $\s$ by $\se$ everywhere, thus \bea \se[t_1,t_2] & = &
B_+^{r(t_1)}(\se[\tilde{u}(B_-(t_1)),t_2])+B_+^{r(t_2)}(\se[t_1,\tilde{u}(B_-(t_2))])\nonumber\\
 & & +B_+^{C[r(t_1),r(t_2)]}(\se[\tilde{u}(B_-(t_1)),\tilde{u}(B_-(t_2))]), \eea
 $\tilde{u}(\prod_{i=1}^k
t_i)=\se[t_1,\tilde{u}(\prod_{i=2}^k t_i)]$, $\tilde{u}(t_i)=t_i$.
The algebras so obtained are commutative algebras if $C$ fufills
the requirements listed above. Commutativity follows immediately
by induction over the number of vertices. Let us prove, again by
induction, associativity in some detail to get acquainted with the
use of $B_+^x,B_-$. Our proof is essentially an elaborated version
of the one in \cite{MEH}. Let $T_1,T_2,T_3\in {\cal H}(A)$. For
$\se[T_1,\se[T_2,T_3]]$ we find (the action of $\tilde{u}$ is
implicitly understood in any appearance of $B_-$) {\small \bea
\se[T_1,\se[T_2,T_3]]
 & = &
 \se[\left(B_+^{r(T_1)}(\se[B_-(T_1),T_2])+
 B_+^{r(T_2)}(\se[T_1,B_-(T_2)])\right.\nonumber\\
 & &
\left.+B_+^{C[r(T_1),r(T_2)]}(\se[B_-(T_1),B_-(T_2)])\right),T_3]\nonumber\\
& = & B_+^{r(T_1)}\left[\se[\se[B_-(T_1),T_2],T_3]
\right]\label{eq7}\\ & & +
B_+^{r(T_3)}\left[\se[B_+^{r(T_1)}(\se[B_-(T_1),T_2]),B_-(T_3)]
\right]\label{eq8}\\ & & +
B_+^{C(r(T_1),r(T_3))}\left[\se[\se[B_-(T_1),T_2],B_-(T_3)]
\right]\label{eq9}\\ & & +
B_+^{r(T_2)}\left[\se[\se[T_1,B_-(T_2)],T_3] \right]\label{eq10}\\
& & +
B_+^{r(T_3)}\left[\se[B_+^{r(T_2)}(\se[T_1,B_-(T_2)]),B_-(T_3)]
\right]\label{eq11}\\ & & +
B_+^{C(r(T_2),r(T_3))}\left[\se[\se[T_1,B_-(T_2)],B_-(T_3)]
\right]\label{eq12}\\ & & +
B_+^{C(r(T_1),r(T_2))}\left[\se[\se[B_-(T_1),B_-(T_2)],T_3]
\right]\label{eq13}\\ & & +
B_+^{r(T_3)}\left[\se[B_+^{C(r(T_1),r(T_2))}\se[B_-(T_1),B_-(T_2)],B_-(T_3)]
\right]\label{eq14}\\ & & +
B_+^{C(C(r(T_1),r(T_2)),r(T_3))}\left[\se[\se[B_-(T_1),B_-(T_2)],B_-(T_3)]
\right],\label{eq15} \eea} while for $\se[\se[T_1,T_2],T_3]$ we
find {\small \bea \se[\se[T_1,T_2],T_3]
 & = &
 \se[\left(B_+^{r(T_1)}(\se[B_-(T_1),T_2])+B_+^{r(T_2)}(\se[T_1,B_-(T_2)])\right.\nonumber\\
 & &
\left.+B_+^{C[r(T_1),r(T_2)]}(\se[B_-(T_1),B_-(T_2)])\right),T_3]\nonumber\\
& = &
B_+^{r(T_1)}\left[\se[B_-(T_1),B_+^{r(T_2)}(\se[B_-(T_2),T_3])]
\right]\label{eq16}\\ & &
+B_+^{r(T_2)}\left[\se[T_1,\se[B_-(T_2),T_3]]
\right]\label{eq17}\\ & &
+B_+^{C(r(T_1),r(T_2))}\left[\se[B_-(T_1),\se[B_-(T_2),T_3]]
\right]\label{eq18}\\ & &
+B_+^{r(T_1)}\left[\se[B_-(T_1),B_+^{r(T_3)}(\se[T_2,B_-(T_3)])]
\right]\label{eq19}\\ & &
+B_+^{r(T_3)}\left[\se[T_1,\se[T_2,B_-(T_3)]]
\right]\label{eq20}\\ & &
+B_+^{C(r(T_1),r(T_3))}\left[\se[B_-(T_1),\se[T_2,B_-(T_3)]]
\right]\label{eq21}\\ & &
+B_+^{r(T_1)}\left[\se[B_-(T_1),B_+^{C(r(T_2),r(T_3))}\se[B_-(T_2),B_-(T_3)]]
\right]\label{eq22}\\ & &
+B_+^{C(r(T_2),r(T_3))}\left[\se[T_1,\se[B_-(T_2),B_-(T_3)]]
\right]\label{eq23}\\ & &
+B_+^{C(r(T_1),C(r(T_2),r(T_3)))}\left[\se[B_-(T_1),\se[B_-(T_2),B_-(T_3)]]
\right]. \label{eq24}\eea} We assume associativity at $n$ vertices
to find (\ref{eq7}) $ = $ (\ref{eq16}) $+$ (\ref{eq17}) $+$
(\ref{eq18}), (\ref{eq8}) $+$ (\ref{eq11}) $+$ (\ref{eq14}) $ =$
(\ref{eq20}), (\ref{eq10}) $=$ (\ref{eq17}), (\ref{eq9}) $ =$
(\ref{eq21}), (\ref{eq12}) $=$ (\ref{eq23}) and (\ref{eq13}) $ =$
(\ref{eq18}). Hence, assuming associativity of $\se$ for products
of trees with up to $n$ vertices, we obtain associativity at $n+1$
vertices provided the map $C$ fulfills the requirements listed
above, which finally ensures (\ref{eq15}) $=$ (\ref{eq24}). The
start of the induction is easily proved with such a map $C$.

In the next section, we will consider an alphabet essentially
given by an infinite set of skeleton graphs in Yukawa theory,
graded by the loop number, and investigate to what extent we can
formulate a quasi-shuffle identity. Hence we have to investigate
if we can find an appropriate  map $C$.

The following result shows that that the transition from a shuffle
identity to a quasi-shuffle identity essentially measures the
non-vanishing of the commutator $[B_+,B_-]=B_+B_--B_-B_+$.  We
have
\be
u([B_+,B_-](X))=0. \ee Proof: Proceed by induction on the number
of vertices, using the definitions of $\s$ and $B_+B_-$.

\noindent Figure (\ref{F1}) gives some $\se$-products.
\bookfig{}{F1}{F1}{The first row gives the $\se$-product
$\se[t_1(a),t_1(b)]$ of two decorated roots. The second row gives
the product $\se[T_{3_2}(c,a,b),t_1(d)]$ of a rooted tree, 'the
claw' $t_{3_2}$, with three vertices, both connected to the root,
which is decorated by the letter $c$ while the other two vertices
are decorated by letters $a,b$, with a root decorated by the
letter $d$.}{4.5}
\section{Yukawa Theory}
In this first approach towards quasi-shuffle identities in quantum
field theories, we will study vertex corrections at zero momentum
transfer in massless Yukawa theory. Hence,  let us consider
two-line irreducible subdivergence-free four-fermion skeleton
graphs $K^{[i]}$ in massless Yukawa theory. Upon closure, we
obtain an overall divergent subdivergence-free vertex correction
corresponding to a graph $\Gamma^{[i]}$, cf.~Fig.(\ref{F2}), at
zero momentum transfer. At each loop number, we have a finite
number of such primitive graphs, and the graphs $K^{[i]}$ provide
an example of an alphabet $A$ in the sense above.
\bookfig{}{F2}{F2}{ Kernels and vertex corrections. We give
kernels $U(K^{[i]})(k,q)$, $i=1,2,3,4$. The momenta $k$ and $q$
enter and leave the diagrams on the sides indicated. We also give
the vertex correction $U(\Gamma^{[2,3]})(q)$.}{5}

To each graph $K^{[i]}$ belongs an analytic expression
$U(K^{[i]})(k,q)$ and the closure to a vertex correction is
defined by the integral \be U(\Gamma^{[i]})(q)=g\int d^Dk
\frac{1}{k^2}U(K^{[i]})(k,q), \ee where $g$ is the renormalized
coupling constant.\footnote{Obviously, the kernels $U(K^{[i]})$
are themselves of order $[g^2]^{1+grad(K^{[i]})}$, where $grad$
gives the loop number.} These graphs $K^{[i]}$ provide an
alphabet, and the vertex correction corresponding to the word
$w=i_1i_2\ldots i_k$ is \bea U(\Gamma^{[w]})(q) & = &
g\int\ldots\int d^Dk_1 \frac{d^Dk_1}{k_1^2}\ldots
\frac{d^Dk_k}{k_k^2}\nonumber\\
 & & \times\;U(K^{[i_1]})(k_1,k_2)\ldots
U(K^{[i_k]})(k_k,q). \eea We can regard $U$ as a character from
the Hopf algebra ${\cal H}(A)$ to the ring of meromorphic
functions ${\cal A}$ as in \cite{CK4,CK3} and the minimal
subtracted counterterm is given by the character $U_-$ of the
unique Birkhoff decomposition $U=U_-^{-1}\star U_+$, where $U_+$
is the character which assigns the renormalized Feynman graph
$U_+(\Gamma^{[w]})$ to the bare unrenormalized $U(\Gamma{[w]})$
\cite{CK4,CK3}.

Consider a different character $\tilde{U}$ defined by
\be
\tilde{U}(\Gamma^{[i]})(q)=g\int d^Dk
\frac{1}{k^2}U(K^{[i]})(k-q,0), \ee so that \bea {\tilde
U}(\Gamma^{[w]})(q) & = & g\int\ldots\int d^Dk_1
\frac{d^Dk_1}{k_1^2}\ldots \frac{d^Dk_k}{k_k^2}\nonumber\\
 & & \times\;{\tilde U}(K^{[i_1]})(k_1-k_2)\ldots
{\tilde U}(K^{[i_k]})(k_k-q), \eea where
$\tilde{U}(K^{[i]})(r)=U(K^{[i]})(r,0)$.

Using the convolution of characters and the antipode $S$ of ${\cal
H}(A)$ , we have $U=\tilde{U}\star \tilde{U}\circ S\star U$ as a
trivial identity. The character $\tilde{U}\circ S\star U$ maps to
functions holomorphic at $D=4$. Indeed, it delivers finite
renormalized Green functions by construction: all divergent
forests in $U(\Gamma^{[w]})$ are compensated by appropriate
forests in $\tilde{U}\circ S(\Gamma^{[w]})$, as such ratios
$\tilde{U}\circ S\star U$ reproduce the forest formula
\cite{CK1,CK4,Khopf}.

We have thus isolated the UV-divergences of such vertex
corrections in $\tilde{U}(\Gamma^{[w]})$, and hence in the minimal
subtracted counterterms $\tilde{U}_-(\Gamma^{[w]})$. Obviously, a
finite rerenormalization  by $U_-\star\tilde{U}_-\circ S$ connects
$U_-$ and $\tilde{U}_-$:
\be
U_-=U_-\star\tilde{U}_-\circ S\star \tilde{U}_-.\ee The character
$\tilde{U}$ can be calculated with considerable ease, as it relies
solely on knowledge of $\tilde{U}(K^{[i]})(q)$, which, being a
function of the single parameter $q^2$, has the form, with
$z:=(D-4)/2$,
\be
\tilde{U}(K^{[i]})(q) =[q^2]^{-1+zgrad(K^{[i]})}F^{[i]}(z),\ee
where $F^{(i]}(z)$ is holomorphic at $D=4$.  Hence,
\be
\tilde{U}(\Gamma^{[i]})(q)=
G_{q^2}(1,1+(grad(K^{[i]})+1)z)F^{[i]}(z), \ee where we use the
$G$-function
\be
G_{q^2}(a,b):= \int d^Dk\frac{1}{[k^2]^a[(k+q)^2]^b}\ee which
evaluates to {\small \be
G_{q^2}(a,b)=[q^2]^{2-z-a-b}\frac{\Gamma(a+b-2+z)\Gamma(2-z-a)\Gamma(2-z-b)}{\Gamma(a)\Gamma(b)\Gamma(D-a-b)}.
\ee} The evaluation of words $w$ via $\tilde{U}(\Gamma^{[w]})(q)$
then delivers products of $G$- and $F^{[i]}$-functions. The
leading pole term of Feynman diagrams fulfills a shuffle identity
\cite{KD}, and indeed, here we find, for example,
\be
\tilde{U}(\Gamma^{[\s[i_1,i_2]})-
\tilde{U}(\Gamma^{[i_1]})\tilde{U}(\Gamma^{[i_2]})\sim
\frac{1}{z}, \ee where $\s[i_1,i_2]=i_1i_2+i_2i_1$,
in this notation. We want to
investigate if the term $\sim 1/z$ can be understood as resulting
from a quasi-shuffle identity. This is possible as a quasi shuffle
identity would precisely demand that this term is of the form
$\tilde{U}(\Gamma^{[C(i_1,i_2)]})$, where $C(i_1,i_2)$ must be
regarded as a single letter, hence be given by an expression
without subdivergences, hence $\sim 1/z$. More precisely,
we want to know if, for
arbitrary letters represented as
$\tilde{U}(K^{[i_1]}),\tilde{U}(K^{[i_2]})$, we can define an
analytic expression $\tilde{U}(K^{[C(i_1,i_2)]})$ such that
\be
\tilde{U}(\Gamma^{[\se[i_1,i_2]})-
\tilde{U}(\Gamma^{[i_1]})\tilde{U}(\Gamma^{[i_2]})
\label{check}\ee is finite. Here,
\be
\tilde{U}(\Gamma^{[\se[i_1,i_2]})=
\tilde{U}(\Gamma^{[i_1i_2]})+\tilde{U}(\Gamma^{[i_2i_1]})+\tilde{U}(\Gamma^{[C(i_1,i_2)}).
\ee Let us now define such a map which assigns to any two letters
a new one. We define
\be
\tilde{U}(K^{[C_2(i_1,i_2)]})(p):=\int d^Dr
\tilde{U}(K^{[i_1]})(r)\frac{1}{r^2}\tilde{U}(K^{[i_2]})(r+p)\frac{(r+p)\cdot
r}{(r+p)^2}.\ee This map $C_2$ crosses kernels and is best
explained in Fig.(\ref{F3}).
\bookfig{}{F3}{F3}{$\tilde{U}(K^{[i_j]})(q)=[q^2]^{[-1-zn_j]}F^{[i_j]}(z)$
so that the map $C_2$ delivers a crossing of these kernels by
essentially raising the crossed bosonic propagators to appropriate
powers,
$\tilde{U}(K^{[C_2(i_1,i_2)]})(p)=[p^2]^{-1+z(n_1+n_2+1)}F^{[C_2(i_1,i_2)]}(z)
$.}{5}

It evaluates to \bea \tilde{U}(K^{[C_2(i_1,i_2)]})(p) & = &
[p^2]^{-1+z(1+n_1+n_2)} F^{[i_1]}(z)F^{[i_2]}(z)\nonumber\\ & &
\times\; [G_1(2+n_1z,2+n_2z)-G_1(1+n_1z,2+n_2z)\nonumber\\
 & & -G_1(2+n_1z,1+n_2z)],
\eea where $n_j=deg(K^{[i_j]})$, $j=1,2$. Closure to a vertex
correction delivers \bea \tilde{U}(\Gamma^{[C_2(i_1,i_2)]})(q) & =
& G_{q^2}(1,1+z(2+n1+n2)) F^{[i_1]}(z)F^{[i_2]}(z)\nonumber\\
 & & \times[G_1(2+n_1z,2+n_2z)\nonumber\\
  & & -G_1(1+n_1z,2+n_2z)-G_1(2+n_1z,1+n_2z)].\eea
Explicit evaluation, ie.~expansion in $z$, shows that the limit
\be \lim_{z\to 0}
[G_1(2+n_1z,2+n_2z)-G_1(1+n_1z,2+n_2z)-G_1(2+n_1z,1+n_2z)] \ee
exists  and that Eq.(\ref{check})) is fulfilled, solely due to the
remarkable properties of $G$-functions. The $F{[i]}$-functions
play no role here as our definition of $\tilde{U}$ reduces their
role to mere coefficients. Also, one immediately confirms that
$C_2(i_1,i_2)=C_2(i_2,i_1)$.

To check associativity of $C_2$ let us define
\be
J(n_1,n_2):=G_1(2+n_1z,2+n_2z)-G_1(1+n_1z,2+n_2z)-G_1(2+n_1z,1+n_2z).
\ee Then, one finds
\be
J(n_1,n_2)J(n_1+n_2,n_3)-J(n_2,n_3)J(n_1,n_2+n_3)\sim z, \ee which
suffices to show that
\be
\tilde{U}(\Gamma^{[C_2(C_2(i_1,i_2),i_3)]})(q)-\tilde{U}(\Gamma^{[C_2(i_1,C_2(i_2,i_3))]})(q)
\ee is finite.

So far we worked up to finite parts. The results  show that we
identify a shuffle algebra in the UV-divergences of iterated
vertex corrections, and we realized the beginning of a shuffle
identity in Eq.(\ref{check}). The finite parts which we
disregarded so far will be essential in the full quasi-shuffle
algebra. They will appear inside further divergent loop
integrations, and hence modify the latter in a non-trivial manner.

Let us finish this paper with an investigation to what extent
these finite parts obstruct a quasi-shuffle identity.

 The violation of the
quasi-shuffle identity in the finite part contributes, when
integrated against a further letter $a$, by
\[
C_3(a,i_1,i_2)= \int \frac{d^Dk}{k^2}\left[
\tilde{U}(\Gamma^{[B_+,B_-](i_1,i_2)})(k)-\tilde{U}(\Gamma^{C_2(i_1,i_2)})(k)\right]K^{[a]}(k-q,0).
\] Here, $$
\Gamma^{[B_+,B_-](i_1,i_2)}=\Gamma^{[i_1i_2]}+\Gamma^{[i_2i_1]}-\Gamma^{[i_1]}
\Gamma^{[i_2]}.$$ This has no contribution to the violation of
associativity, as it is demonstrated in Fig.(\ref{F4}). It can be
incorporated in a redefinition of $C_2$ respecting the
quasi-shuffle algebra. \bookfig{}{F4}{F4}{$\se[\se[a,b],c]$
given explicitly. From the three terms in the rectangle we read
off the contributions of $C_3$. Those three terms provide
decorations, in which $C_2$ terms (abbreviated by $C$ in the
figure) are nested. $C_3$ terms contributes to $\se[\se[a,b],c]$in
the cyclic sum $C_3(a,b,c)+C_3(b,c,a)+C_3(c,a,b)$. As $C_3$ is
commutative in the last two arguments, $C_3(a,b,c)=C_3(a,c,b)$, no
violations of associativity arise from $C_3$, as $\se[a,\se[b,c]]$
provides the same terms.}{5}

More interesting is the finite violation of associativity
generated by $C_2$. Explicitly, one finds
\be
J(n_1,n_2)J(n_1+n_2,n_3)-J(n_2,n_3)J(n_1,n_2+n_3)=-4z(n_1-n_3)+{\cal
O}(z^2) \ee so that the violation of associativity is proportional
to the difference between the loop numbers, the degree,  of the
first and last letter. Hence, the finite terms violate
associativity in a well-defined manner: the pentagon equation
\[
(n_1-n_3)+(n_1-n_4)+(n_2-n_4)=2n_1+n_2-n_3-2n_4=(n_1+n_2-n_4)+(n_1-n_3-n_4)
\] is satisfied, cf.~Fig.(\ref{F5}). \bookfig{}{F5}{F5}{The
pentagon. Bracket configurations $(1(23))4$ correspond to
$\se(\se(i_1,\se(i_2,i_3)),i_4)$, $grad(K^{[i]})=:n_i$.}{4}
\section{Conclusions}
We have shown how quasi-shuffle algebras can be naturally
formulated using the operators $B_+,B_-$. We then investigated
representations of this algebra by Feynman diagrams, and found
that, in the case of Yukawa theory considered here, such
representations could be obtained modulo finite violations of
associativity, and that these representations deviate by finite
parts from a quasi-shuffle identity. These violations obey a
pentagon identity in the finite part, and it will be an
interesting exercise to work out the coherence laws of higher
orders in $z$ in the future. Also, the generalization to other
theories and Green functions is possible using appropriate
representation of $B_\pm$ on Feynman graphs and will be presented
in future work. We regard the results in this paper as a first
attempt towards an algebraic understanding of the number-theoretic
content of Feynman diagrams.
\section*{Acknowledgements}
This work was done for the Clay Mathematics Institute during a
stay at Harvard University. It is a pleasure to thank Olivier
Grandjean, Arthur Jaffe and David Kazhdan for stimulating
discussions during that stay. Also, I like to thank Pierre Cartier
for the invitation to participate in his stimulating seminar on
Euler-Zagier sums at the Institut Henri Poincar\'e and thank David
Broadhurst and Alain Connes for an ongoing collaboration which
enabled these results.

\end{document}